\documentclass{aa}
\usepackage{graphicx}
\usepackage{xcolor}
\usepackage{txfonts}
\usepackage{amsmath}

\newcommand{\Fig}[1]{Figure~\ref{#1}}

\newcommand{\curl}{ {\bf \nabla} \times}

\newcommand{\Figss}[2]{Figures~\ref{#1}--\ref{#2}}
\newcommand{\Eq}[1]{Equation~(\ref{#1})}

\newcommand{\Tab}[1]{Table~\ref{#1}}

\def\bl{Babcock--Leighton}

\newcommand{\mps}{m~s$^{-1}$}

\newcommand{\cmss}{cm$^2$~s$^{-1}$}

\newcommand{\etasurf}{\eta_{\mathrm{surf}}}
\def\Rs{R_{\odot}}
\newcommand{\vect}[1]{\boldsymbol{#1}}

\authorrunning{P. Kumar et al.}
\titlerunning{Short-term periods in \bl\ dynamo}
\usepackage[switch]{lineno}  

\begin{document}


\title{Quasi-Biennial Oscillations and Rieger-type Periodicities in a \bl\ Solar Dynamo}


\author{Pawan Kumar \inst{1,2} \ Belur Ravindra \inst{2} \
Partha Chowdhury \inst{3} \and Bidya Binay Karak \inst{4}
          }

\institute{Physics and Astronomy Department, University of Florence (FI), 50019, Italy\\
\email{pawan.kumar@unifi.it}
\and
Indian Institute of Astrophysics, Koramangala, Bengaluru, 560034, India\\
\email{ravindra@iiap.res.in}
\and
University College of Science and Technology, Department of Chemical Technology, University of Calcutta, 92, A.P.C. Road, Kolkata, 700009, West Bengal, India\\
\email{parthares@gmail.com}
\and
Department of Physics, Indian Institute of Technology (BHU) Varanasi 221005, India\\
\email{karak.phy@iitbhu.ac.in}
             }


 
  \abstract
{The Sun's magnetic field shows the 11-year solar cycle and shorter periodicities, popularly known as the quasi-biennial oscillations (QBOs) and Rieger-type periods, or ``season of the Sun." Although several theories have been proposed to explain the origin of QBOs and Rieger-type periods, no single theory has widespread acceptance. }
{We explore whether the \bl\ dynamo can produce Rieger-type periodicity and QBOs and investigate their underlying physical mechanisms. }
{We use the observationally guided three-dimensional kinematic \bl\ dynamo model, which has emerged as a successful model for reproducing many characteristic features of the solar cycle. We use Morlet wavelet and global wavelet power spectrum techniques to analyze the data obtained from the model.}
{In our model, we report QBOs and Rieger-type periods for the first time. Further, we investigate the individual \bl\ parameters (fluctuations in flux, latitude, time delay and tilt scatter) role in the occurrence of QBOs and Rieger-type periods. We find that while fluctuations in the individual parameters of the \bl\ process can produce QBOs and Rieger-type periodicity, their occurrence probability is enhanced when we consider combined fluctuations of all parameters in the \bl\ process. Finally, we find that with the increase of dynamo supercriticality, the model tends to suppress the generation of Rieger-type periodicity. Thus, this result supports earlier studies that suggest the solar dynamo is not highly supercritical.}
{}

\keywords{Sun: activity -- magnetic fields -- sunspots --solar dynamo -- Sun: oscillations -- Methods: data analysis }

\maketitle
%

\section{Introduction}
\label{sec:intro}
The global solar magnetic field shows cyclic variation in its activity levels with an average period of 11 years (or 22 years when considering the polarity). Beyond this 11-year oscillation, there are long-term modulations, which are
best seen in the sunspot number or its proxies \citep[$^{10}$Be and $^{14}$C;][]{Uso23}, e.g., $\sim$ 90 years Gleisberge cycle \citep{glessberg}, $\sim$ 210 years Suess cycle \citep{Suess}, and Grand minimum \citep[like the Maunder minimum;][]{Eddy}.
Besides these long-term modulations and variations in periodicities and amplitude, the solar cycle shows Rieger-type periodic variations ($<$ 1-year), which are so-called bursts of activity or seasons of the Sun \citep{Rieger, Scott15}. 
In addition to these Rieger-type periodicities, several mid-range periodicities have
been observed between 1 and 11 years, known as quasi-biennial oscillations \citep[QBOs;][]{Baz00,RZ08, AV09, Chowdhury22, RB22}. 
Studies suggest that QBOs attain their highest amplitude during solar maxima and become weaker during the minimum phase of the solar cycle; thus, the amplitude of the QBO's signals is in phase with the 11-year solar cycle. 
Moreover, QBOs develop in both hemispheres independently with variable periodicity \citep{Baz14, KJ23}
and show asymmetry, intermittent behavior, and presence in only a few hemispheric cycles of sunspot number \citep{RB22}.

There is extensive literature available on the origin of QBOs and Regier-type periodicities in solar activity. However, the conclusions of different authors on the origin of these periodicities are different. The majority of authors suggest that QBOs arise intrinsically from the solar dynamo process, which itself drives the 11-year solar cycle \citep[e.g.,][]{Benevo98, Howe00, KS05, RZ08, Cho14, IF19}.
There are some well-known explanations that have been given for Reiger-type periodicity and QBOs, for example, two-dynamo process, in which one operates near solar surface and the other at the base of the convection zone \citep{Kap16};  instabilities of magnetic Rossby waves in the solar tachocline \citep{PC09, Za10, MK23};   beating between dipolar and quadrupolar magnetic field configurations generated by the solar dynamo \citep{RS13}; the typical lifetimes of complex active regions \citep{PC13}.
Moreover, \citet{Dik18, Dikpati21} suggested that the non-linear oscillations in the tachocline might be responsible for the emergence of QBOs and Regier-type periodicities. 
They also provided a quantitative physical mechanism for forecasting the strength and duration of the bursty seasons or seasons of the Sun several months in advance \citep{Dik17}.
Furthermore, the results of MHD numerical simulations \citep{Kar15, Kap16, Strugarek18} also show the existence of two-cycle modes with longer and shorter periods, and people believe that this may also be a possible candidate for an explanation of Rieger-type periodicities.

The solar magnetic field and its cycle-to-cycle variations are believed to be the result of the dynamo process, which operates in the solar convection zone (SCZ). Dynamo is a cyclic process in which toroidal and poloidal fields support each other.  In the classical $\alpha\Omega$-type dynamo, the
helical $\alpha$ generates the poloidal field from the toroidal field, while the $\Omega$ effect produces the toroidal field from the poloidal one through differential rotation \citep{Pa55, SKR66}. 
Under certain conditions, this cyclic process continues and sustains the long-term evolution of the solar magnetic field.
However, in recent years, observational studies \citep{Das10, KO11, Mord22, CS23} suggest that the decay and dispersal of tilted bipolar magnetic regions (BMRs) on the solar surface generate the poloidal field, which is popularly known as the \bl\ mechanism \citep{Ba61, Leighton69}. 
In recent decades, \bl-type dynamo models, which involve the \bl\ mechanism in solar polar field generation, have emerged as a popular paradigm for explaining various key features of the observed solar magnetic field, including polarity reversals and double peaks \citep{Gnev77, KMB18, Mord22}; solar cycle prediction \citep{HC19, Petrovay20, Kumar21, Kumar22}; poleward migration of surface magnetic field; equatorward migration of the toroidal field; and solar cycle variability \citep{Kar14a, Cha20, Karak23, Kumar24}. 
The key strength of \bl\ models is that it has strong observational support for the toroidal-to-poloidal field conversion part of the model, and in the model, we can include fluctuations in BMR properties, which are observable and quantifiable \citep{JCS14, CS23, Sreedevi24}. These characteristics enable the models to reproduce and explain various observed and irregular features of the solar magnetic cycles.

Although the \bl\ dynamo model successfully reproduces many observed features of the solar magnetic field, some characteristics remain unexplained. For example, \citet{IF19} found that a two-dimensional (2D) turbulent $\alpha$ effect dynamo model is capable of generating quasi-biennial oscillations (QBOs); however, the 2D \bl\ dynamo model fails to reproduces these features. They suggested that in the turbulent $\alpha$ dynamo model, the Lorentz force provides a feedback mechanism on the flow fields, which enables the model to generate QBOs.

In this work, we explore whether the \bl\ dynamo can produce Rieger-type periodicity and QBOs and the underlying physical mechanism.
We use a three-dimensional kinematic \bl\ dynamo model, STABLE  \citep[Surface Flux Transport And Babcock-LEighton;][]{MD14}. 
The STABLE model captures the BMR properties more realistically and makes a close connection between the model and observation. In the \bl\ process, the poloidal field is generated through the decay and dispersal of tilted BMRs. Therefore, BMR properties are crucial in the \bl\ mechanism.
We shall focus on the various fluctuating parameters in the \bl\ process  that are caused by turbulent convection \citep{Kumar24}.  
Specifically, we investigate the roles of four key BMR properties, 
such as fluctuations in (i) the tilt angle, (ii) flux, (iii) time delay of emergence, and (iv) emergence latitude.

The structure of this paper is as follows. In Section \ref{sec:model}, we give the description and implementation of our model used to generate the sunspot cycle and surface magnetic field. Section \ref{sec:methods} outlines the methodology adopted to analyze the model data, and in Section \ref{sec:results}, we present and discuss the results obtained from the dynamo model. The last section presents the conclusions of the study.

\section{Model}
\label{sec:model}
To understand and demonstrate QBOs and Rieger-type periods, we perform our study using
three-dimensional \bl\ dynamo model STABLE \citep{MT16}. The \bl\ process involves stochastic fluctuations and non-linearity and is responsible for generating the Sun’s poloidal magnetic field through the decay and dispersal of tilted BMRs on the solar surface.
This poloidal field eventually gives the toroidal field through differential rotation and produces new BMRs via magnetic buoyancy, and thus the sunspot cycle. 
In the STABLE model, BMRs are deposited on the surface based on the toroidal field present in the convection zone (CZ). The model's BMR prescriptions are to be specified by observation.
Thus, this model produces observed features of the solar magnetic field reasonably well \citep{MD14, MT16}, including the correct latitude dependence of the polar field --- the higher the latitude of BMR emergence, the lesser the amount of polar field generation \citep{Kumar24}, latitude quenching \citep{Kar20}, polar rush, triple reversal \citep{Mord22}, and irregular cycles and grand minima \citep{KM17, KM18}. The details of the model used in this study are described below.


STABLE dynamo model, which was originally developed by Mark Miesch \citep{MD14, MT16} and improved by \citet{KM17} to make a close connection of BMRs eruptions with observations. 
This model realistically captures the \bl\ process using the available surface observation of BMRs and large-scale flows such as meridional circulation and differential rotation.
In this model, we solve the induction equation in spherical coordinates $(r,\theta,\phi)$ for the whole SCZ.
 
\begin{equation}
\frac{\partial \vect{B}}{\partial t} = \vect{\curl} \left[ (\vect{V} +\vect{\gamma})
\times \vect{B} - \eta_t \vect{\curl} \vect{B} \right],
\label{eq:ind}
\end{equation}
with $0.69 \Rs \le r \le \Rs$, where $\Rs$ is the solar radius, $0\le\theta \le \pi$ and $0\le \phi \le 2\pi$. In this study, the model used is a kinematic \bl\ dynamo in which $\vect{V}$ is the velocity field such that, 
\begin{equation}
\vect{V} = v_r(r,\theta)\hat{r} + v_\theta(r,\theta)\hat{\theta} + r\sin\theta \Omega(r,\theta)\hat{\phi},
\end{equation} 
where $v_r$ and $v_\theta$  are the component of axisymmetric meridional flow and ($\Omega =  v_\phi / r\sin\theta)$ is the differential rotation.
For meridional circulation, we consider the single-cell circulation profile which closely resembles surface observations \citep{RA15, Gizon20} and was used earlier in \citet{KC16, KM17, Kumar24}. Near the surface, we consider meridional flow speed is  of 20 \mps\ towards the pole and near the base of the CZ it is of 2 \mps\ and it smoothly goes to zero at the lower boundary $(0.69 \Rs)$ of CZ. The differential rotation $(\Omega)$ profile used in this model roughly captures the observed properties as inffered by the heliosysmology \citep{Schou98}.

In the Induction \Eq{eq:ind}, $\gamma$ represents the magnetic pumping which helps to suppress the loss of toroidal flux through surface due to diffusivity. 
In this model, we consider a radially downward magnetic pumping of speed 20 \mps\ in the near surface layer of the Sun $(r\ge 0.9\Rs)$; see equation (3) of \citet{KM17}.
Magnetic pumping makes our \bl\ dynamo models consistent with the Surface Flux Transport (SFT) model \citep{CS12, Kumar24}. Moreover, the magnetic pumping helps our model to produce an 11-year magnetic cycle at very high turbulent diffusivity as inferred in the observations (of the order of $10^{12}$ \cmss) \citep{KC16}.
Next, we consider an effective radial-dependent turbulent diffusivity represented by $\eta_t$ in \Eq{eq:ind} such that

\begin{eqnarray*}
\eta_t(r) = \eta_{RZ} + \frac{\eta_{cz}}{2}\left[1 + \mathrm{erf} \left(\frac{r - 0.715\Rs}
{0.0125\Rs}\right) \right] \\ \nonumber
+\frac{\etasurf}{2}\left[1 + \mathrm{erf} \left(\frac{r - 0.956\Rs}
{0.025\Rs}\right) \right],
\label{eq:eta}
\end{eqnarray*}
where $\eta_{RZ} = 10^9$\cmss, $\eta_{surf}= 4.5\times 10^{12}$\cmss, and $\eta_{cz}= 1.5\times 10^{12}$\cmss.

We note that this model does not capture full magnetohydrodynamics in the convection, so the BMRs do not appear automatically. This model has a prescription known as the SpotMaker algorithm for this mechanism \citep{MT16}. First, it computes the spot-producing toroidal field strength near the base of the CZ in a hemisphere presented as,
\begin{equation}
\hat{B}(\theta, \phi, t) = \int_{r_a}^{r_b} h(r) B_{\phi}(r, \theta, \phi, t) \, dr
\label{eq1}
\end{equation}

where $r_a=0.7\Rs$, $r_b=0.715\Rs$ ($\Rs$ is the radius of the Sun) and $h(r ) = h_0 (r - r_a )(r_b - r )$, $h_0$ is a normalization factor.
Further, the model places a BMR on the solar surface when only some conditions are satisfied. First, $\hat{B}(\theta, \phi, t)$ must exceed a critical magnetic field strength value $B_{tc} (\theta)$. The critical magnetic field $B_{tc}$ depends on latitude, making the emergence of BMR difficult at higher latitudes such that:
\begin{equation}
\begin{split}
B_{tc}(\theta) &= B_{t0} \exp \left[ \beta (\theta - \pi/2) \right], ~~~~~~\rm for~~~~ \theta > \pi/2 \\
&= B_{t0} \exp \left[ \beta (\pi/2 - \theta) \right]~~~~~~~~\rm for~~~~\theta\le\pi/2
\label{eq2}
\end{split}
\end{equation}
where $B_{t0}$ is 2 kG and $\beta=5$. This latitude-dependent BMR eruption plays an important role in capturing latitude-dependent quenching \citep{Petrovay20, J20, Kar20} and helps model BMR to be consistent with observations \citep{SWS08, Mandal17}.
Our model produces the first BMR when $\hat{B}(\theta,\phi) > B_{tc}(\theta)$.
Then, after a time dt of the first BMR eruption, the model can produce the next BMR only when these two conditions are satisfied. \\ (i) $\hat{B}(\theta,\phi) > B_{tc}(\theta)$ \\
(ii) $dt\ge \Delta$.\\
Here $\Delta$ is the time delay between two consecutive BMR eruption and follows a log-normal distribution which is obtained from the fitting of observed sunspots and follows as:

\begin{equation}
P(\Delta)= \frac{1} {\sigma_d\Delta \sqrt{2\pi}}{\rm exp}\left[\frac{-(\rm\ln\Delta - \mu_d)^2}{2\sigma_d^2}\right]
\label{eq:delay}
\end{equation}

where $\sigma_d$ and $\mu_d$ are specified as,
$\sigma_d^2=\frac{2}{3}\left[\ln\tau_s - \ln\tau_p\right]$
and
$\mu_d=\sigma_d^2 + \ln\tau_p$.
Here $\Delta$ is the time delay between two successive BMRs (normalized to one day), and $\tau_p = 0.8$ days, $\tau_s = 1.9$ days, which are obtained from the observed solar maxima data. 
Since observation suggests that the time delay of BMR eruption is solar cycle phase-dependent \citep{jiang2018predictability} therefore, we consider $\tau_p$ and $\tau_s$ be magnetic field dependent such that:
\begin{equation}
\tau_p = \frac{2.2} {1 + \left[\frac{B_b}{B_0}\right]^2} \rm days, ~~~~{\rm{and}}~~~~
\tau_s = \frac{20} {1 + \left[\frac{B_b}{B_0}\right]^2} \rm days,
\end{equation}

where,  $B_b$ is the azimuthal averaged toroidal magnetic field in a thin layer that spans from $r = 0.715\Rs$ to $0.73\Rs$ at approximately $15^{\circ}$ latitude and $B_0 =400$ G estimated from observed BMR.
Thus, the delay distribution changes in response to the toroidal field at the base of CZ as the time delay of successive BMR emergence depends on the strength of the toroidal magnetic field in the CZ \citep{Jouve10}. 
After the timing of eruption is decided, we obtained the other properties of the BMR from observations. 
In our model, the flux of BMR specified by the observed log-normal distribution based on the \citet{Mu15} and follows the profile:
\begin{equation}
P(\Phi)=\Phi_0\frac{1}{\sigma_\Phi\Phi\sqrt{2\pi}}{\rm exp}\left[\frac{-(\rm\ln\Phi -\mu_\Phi)^2}{2\sigma_\Phi^2}\right]
\label{eq:flux}
\end{equation}

where, $\Phi_0 = 2$ unless otherwise mentioned, $\mu_\Phi=51.2$, and $\sigma_\Phi=0.77$. We set the magnetic field strength of the BMR to 3 kG at the solar surface.
The radius of the spot is determined automatically on the basis of the flux content obtained from the lognormal distribution. For the spatial separation of the two polarities of BMR, we set the half-distance between the centers of the two spots to be 1.5 times the spot radius \citep{MT16}.
Now, we specify the tilt of the BMRs using Joy's law with Gaussian scatter $\sigma_\delta$ which is taken from observations \citep{SK12, Sreedevi24}. 

\begin{equation}
\delta=\delta_0\cos\theta + \delta_f   
\end{equation}
where $\theta$ is the colatitude and $\delta_0= 35^{\circ}$. $\delta_f$ includes fluctuation in tilt around Joy's law
and follows the Gaussian distribution with $\sigma_\delta \approx15^{\circ}$ \citep[e.g.,][]{Wang15,Sreedevi24}.
As observations demonstrate, for strong cycles the generated polar field is considerably reduced due to tilt quenching \citep{bidisha}.
Therefore, in addition to  latitude quenching as discussed above (\Eq{eq2}), we include a magnetic-field-dependent tilt quenching in the tilt of the form: 1/$\left[1 + (\hat{B}(\theta,\phi, t)/B_0)^2\right]$ (with $B_0=10^5$G), inspired by observations \citep{Jha20, Sreedevi24};
however, see  
\citet{Qin25}, who do not support a dependence of tilt angles on the maximum magnetic field strength of active regions based on the study of mutual validation of SOHO/MDI and SDO/HMI data with the Debrecen Photoheliographic Data (DPD), suggesting further investigation of the tilt data and the other nonlinearities.
Nevertheless, above tilt  nonlinearity helps to stabilize the magnetic field's growth in the \bl\ dynamo and for strong cycles this becomes important as observations suggest \citep{bidisha}.
We note that while the tilt quenching used in our study helps maintain a stable cycle in a highly supercritical regime, the latitude quenching alone can stabilize the dynamo as long as it is not too supercritical \citep{Kar20}.  
For more details of the model, see \citet{KM17}.

\begin{figure*}
\centering
\includegraphics[width=\linewidth]{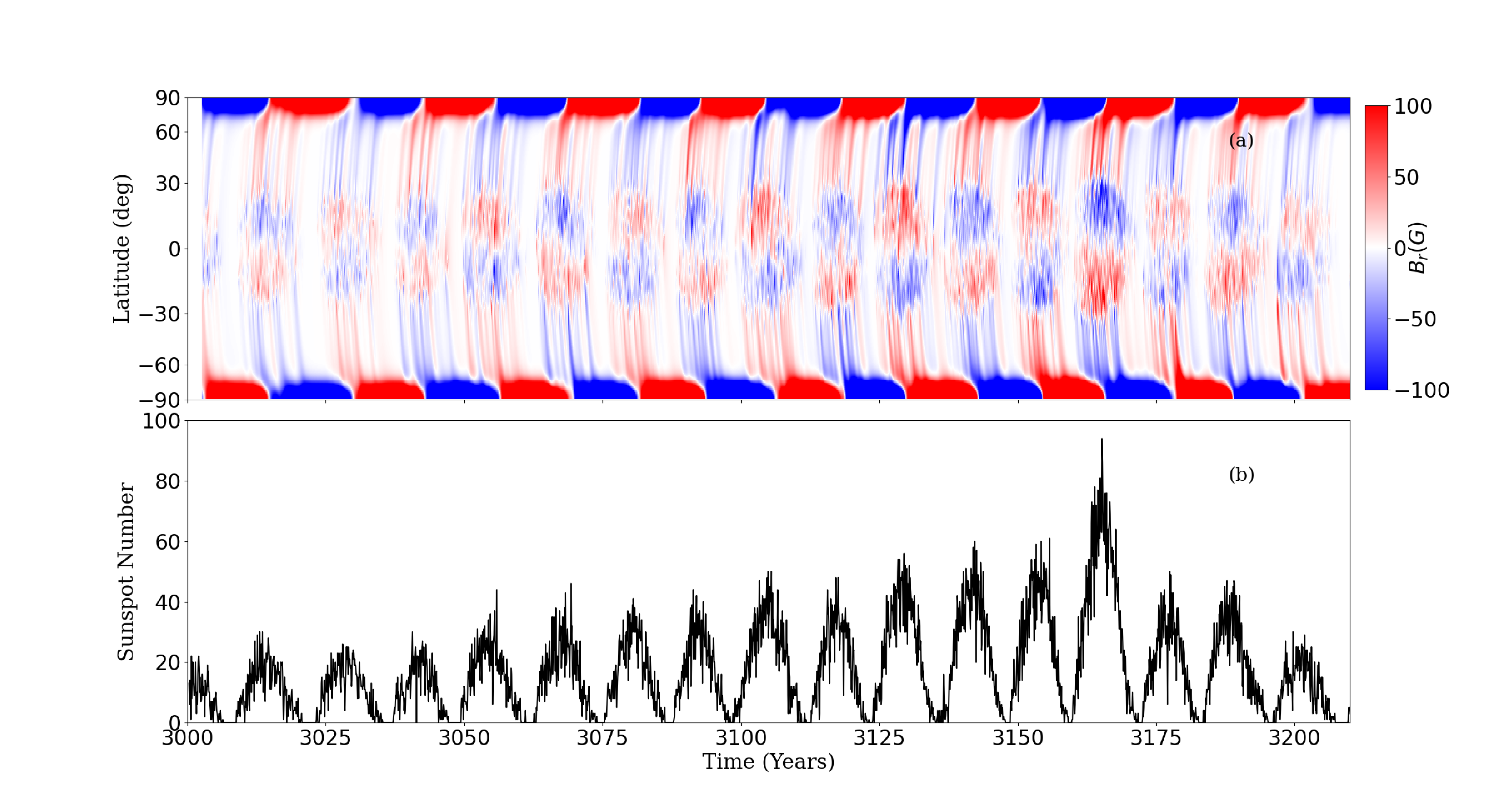}
\caption{Solar cycles obtained from the STABLE dynamo model.
(a) The azimuthally average surface radial field as functions of latitude and time.
(b) Temporal variation of the total (over the full-disk) monthly sunspot number.}
\label{fig:time_series}
\end{figure*}

\section{Methods} \label{sec:methods}

\Fig{fig:time_series}(a) shows the temporal evolution of azimuthally average surface radial field on the solar surface and (b) the time series of the sunspot number obtained from the STABLE dynamo model. 
To analyze Rieger-type periodicity and QBOs signals, we use the bandpass filter technique \citep{Edmonds16}. For data smoothing, we use a Gaussian filter \citep{Hat02} with FWHM = 4 and 8 months for Rieger-type period analysis and FWHM = 12 and 24 months for QBOs. We isolate Rieger-type periodicities and QBOs by subtracting the 8-
month smoothed data from the 4-month smoothed data, and the 24-month smoothed data from the 12-month smoothed data, respectively, following \citet{Baz14}.
Next, we detrend the substracted data and compute the mean and standard deviation to standardize the data.
After data standardization,
we use the wavelet analysis tool \citep{TC98} to study QBOs and Regeir-type periods.  
We use \citet{Morlet82} wavelet function with frequency $\omega_0 =6$ and global wavelet power spectra (GWPS) to detect all periods present in the data set with 95\% significance level. 
The nature of the GWPS power spectrum is similar to the Fourier power spectrum of a time series. We compute the significance level 95\% in the spectra, the same as calculated in the \citet{TC98} and represent it by a red dash line in the GWPS plot.
In all Morlet wavelet spectra, the thin black contours represent periods above the 95\% confidence level, considering the background of red noise. By adopting a strong red-noise background model, we minimize the likelihood of false detections and ensure the reliability of the results. 
A black dashed line shows the cone of inﬂuence (COI) in the Morlet wavelet spectra (\Figss{fig:fig1}{fig:sup} (a) and (c)), which indicates a reduction in wavelet power due to edge effects \citep{TC98}.
The color bars in Figures 2 -- 8 indicate the spectral power range (in $log_{2}$ scale) of the data. 
Moreover, for the reliability and robustness of the results obtained in different cases, we keep the length of the time--series data approximately the same.

\section{Results and Discussion} \label{sec:results}

We compute the monthly sunspot number and flux using the STABLE dynamo model to explore the underlying physical mechanisms and the spatiotemporal variations (QBOs and  Rieger type of periods), as reported in various observational datasets (e.g., sunspot number, sunspot area, subsurface flow fields, etc.) \citep{Baz14, IF21, RB22}.
Previous studies and observations suggest that the BMR tilt angle, magnetic flux of BMR, latitude distribution of BMR, and time delay between consecutive BMR emergence are the primary physical parameters that significantly influence the \bl\ mechanism \citep{KM18, CS23, Kumar24}. These parameters also play a crucial role in the variability of the solar magnetic field and the asymmetry of the solar cycle \citep{Bau04, Kumar24}.
The STABLE model naturally incorporates all these physical parameters and produce cyclic magnetic field.
To investigate whether \bl\ model produces Rieger-type periodicities such as quasi-biennial oscillations (QBOs) and Rieger-type periodicities, we employ Morlet wavelet power spectrum analysis along with the global wavelet power spectrum of both sunspot number and flux data. 
In our model, the BMR's magnetic flux and sunspot number are intrinsically connected, and leads to similar results in both sunspot number and flux analyses. 
Therefore, we present our results based on the sunspot number to maintain consistency with observational data (e.g., sunspots and proxies of the solar magnetic field).

To investigate the physical mechanism of QBO and the Rieger-type periods, we first explore the role of individual fluctuating parameters (tilt scatter, flux variation, and time delay) that influence the \bl\ mechanism. 
We note that to study Rieger-type periodicities and QBOs in
this model, we do not consider the case of the variation in the BMR latitude separately because it is not trivial to vary BMR latitude randomly by hand in
this model.
However, the mean latitude of the BMRs automatically varies with changes in the cycle strength in the model. In addition, we analyze the effect of the fluctuations in the combined parameters.
Finally, in subsection \ref{sec:sup}, we examine the effects of dynamo supercriticality on these periodicities. To do so, we gradually increase the value of $\Phi_0$ within a feasible range for the dynamo, keeping the rest of the parameters the same.

\subsection{Effects of fluctuations in \bl\ process}
In this subsection, we show the results of the individual \bl\ parameters and their combined impact on QBOs and Rieger periodicities. Note that we consider only the full-disk data to analyze individual parameter's effect on these periodicities, as the hemispheric data produce similar results as the full-disk data. 

\subsubsection{Time delay}

\begin{figure}
\includegraphics[width=\linewidth]{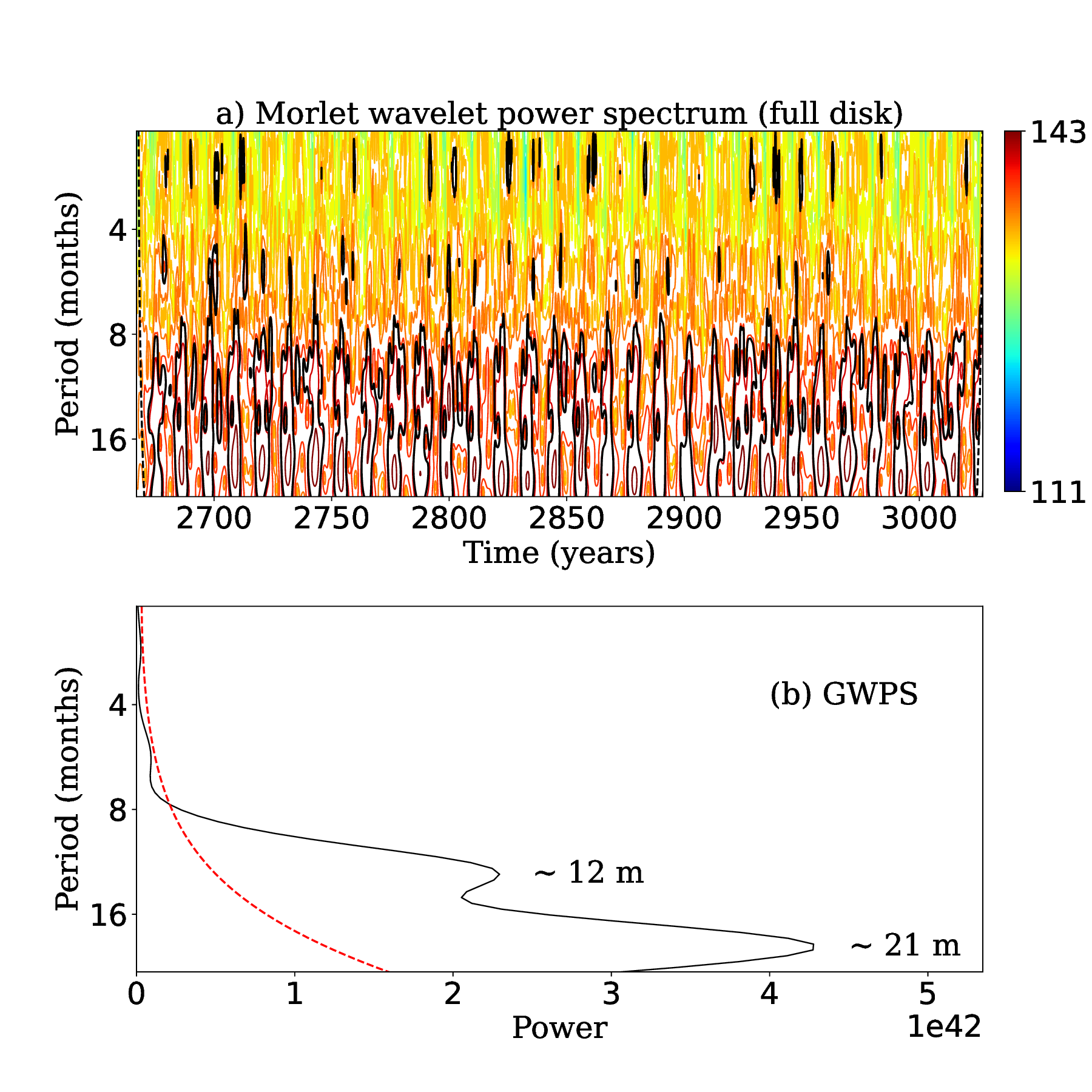}
\includegraphics[width=\linewidth]{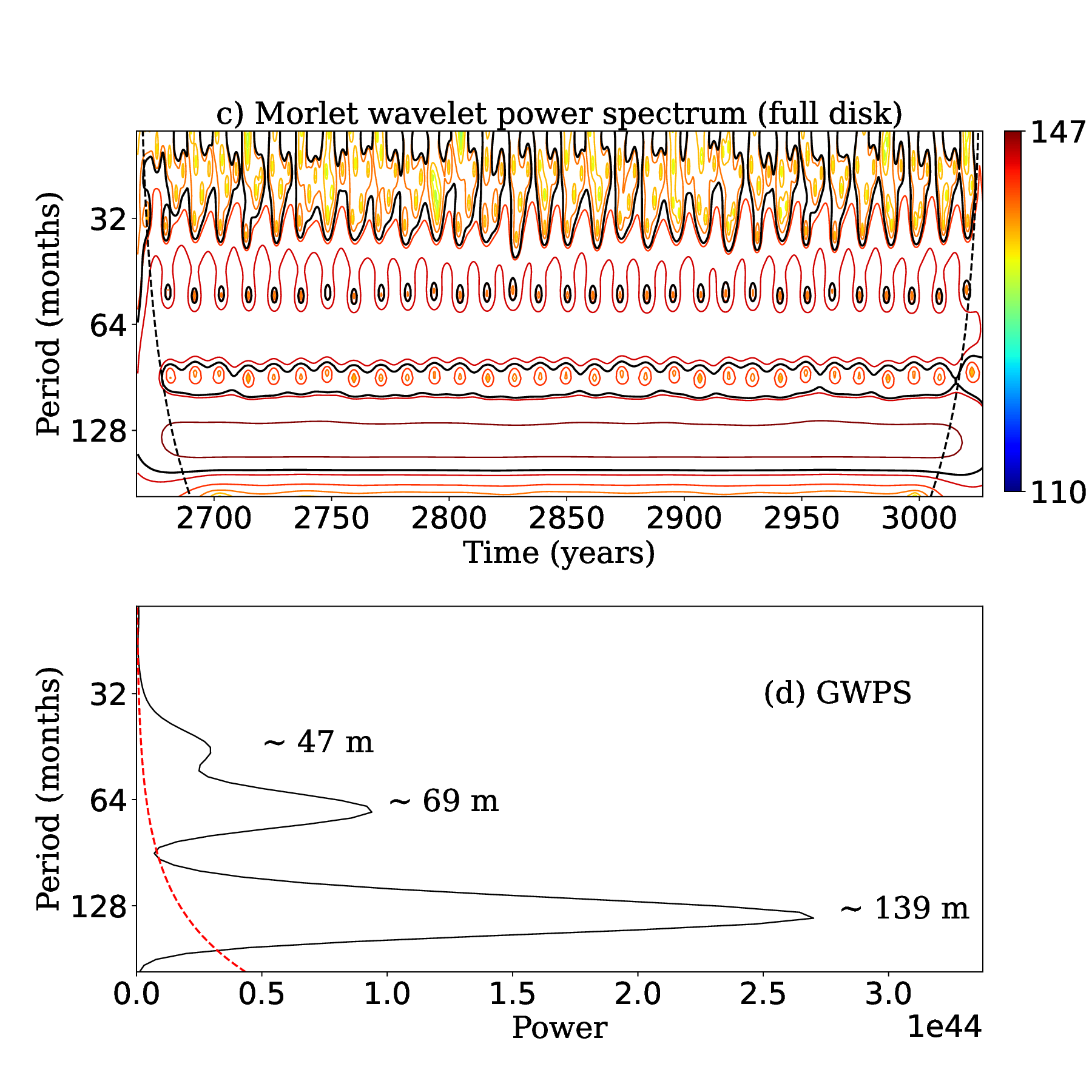}

\caption{
Morlet and global wavelet power spectra of monthly sunspot numbers obtained from the dynamo model with fluctuations in the BMR time delay to study the short-term/Rieger-type periodicities (a-b) and QBOs and other long-term periods including the main cycle period corresponding to the 11-year solar cycle (c-d).
The red dotted lines in all global wavelet power spectra represent the 95\% conﬁdence level and black dotted lines in Morlet wavelet spectra represent COI.
}
\label{fig:fig1}
\end{figure}

First, we consider the effect of time delays in BMR emergence on QBOs and Rieger-type periodicities. To do so, in the STABLE model we set a fixed value of the magnetic flux ($10^{22}$ Mx), set the tilt fluctuations around Joy's law to zero  $(\sigma_\delta=0$), and time delay is computed from the distribution as given by \Eq{eq:delay}. 
We analyze the data using the Morlete wavelet analysis technique and GWPS (see Section \ref{sec:methods}).
\Fig{fig:fig1}(a) and (b) show the analysis of Rieger-type periods and (c) and (d) for QBOs. 
The analysis of the simulated solar cycles reveal prominent periodicities at $\sim$ 12, 21, 47, 69, and 139 months (see \Fig{fig:fig1}). The Morlet wavelet power spectrum and GWPS clearly indicate the presence of QBOs in the data, with 95\% confidence level.
In \Fig{fig:fig1}(a) and (c), the black dotted lines mark the COI in the Morlet wavelet analysis, and black thin lines show periodicity contours above the 95\% confidence level. In contrast, in \Fig{fig:fig1}(b) and (d), the red dotted line denotes the 95\% confidence boundary for GWPS.
The above analysis shows that only QBO signals are present in the data. This result suggests that the stochasticity in time delay of BMR emergence can individually produce QBOs but not Rieger-type periods with 95\% of significance level.
This result is consistent with the findings of \citet{WS03}, who reported quasi-periodic variations with timescales of $\sim$ 1--3 years in the equatorial dipole moment and open flux derived from both surface flux transport (SFT) simulations and observed group sunspot areas. They suggested that the physical mechanism of these short-term periodicities is the stochastic fluctuations in the emergence rate of BMRs and their random longitudinal distribution on the solar surface.

\subsubsection{Flux variation}
\begin{figure}
\includegraphics[width=\linewidth]{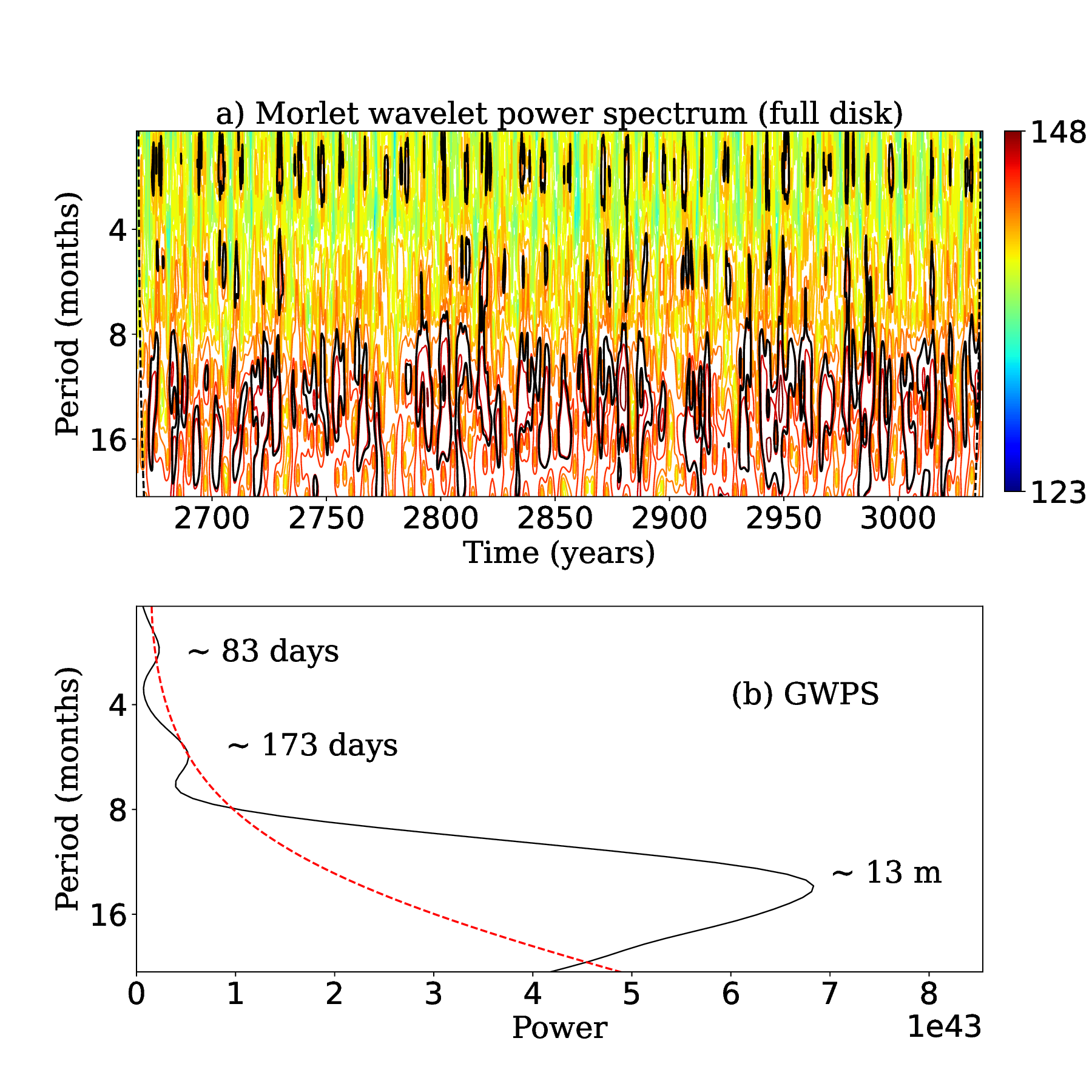}
\includegraphics[width=\linewidth]{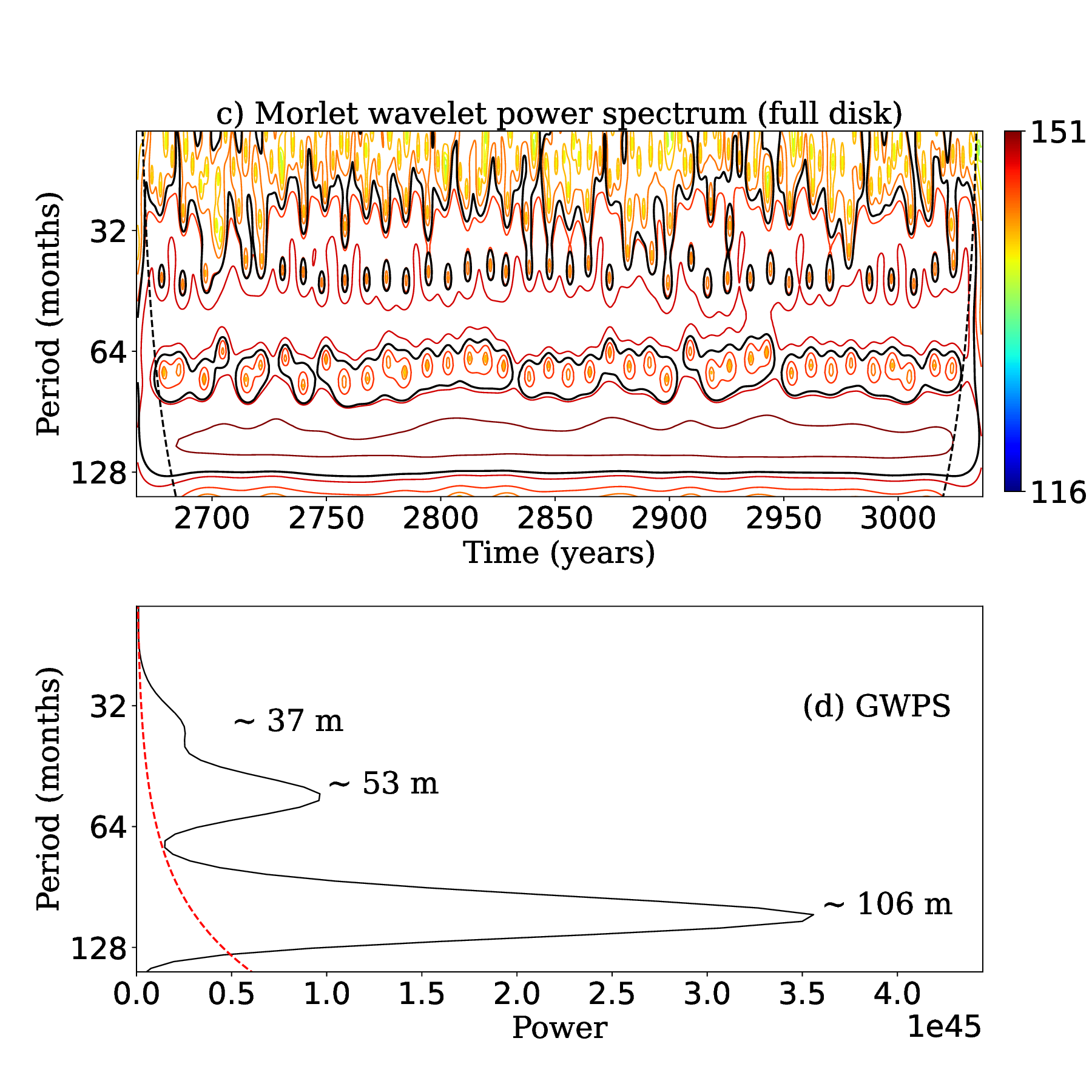}
\caption{Same as \Fig{fig:fig1}, but
from the case in which we consider fluctuation in flux only.
}
\label{fig:fig2}
\end{figure}

Next, we examine the effect of flux variation on the QBOs and Rieger-type periodicity. For this, we do not fix the flux at $10^{22}$~Mx but  computed from the distribution given in \Eq{eq:flux} so that the individual flux varies in a wide range consistent with obsevations. And we  fix the time delay and as before we do not include any scatter scatter around Joy's law. 
Applying Mortlet wavelet analysis and GWPS to the sunspot flux time series obtained from this simulation, we find periodicities of $\sim$ 83 and 173 days, and 13, 37, 53, and 106 months (\Fig{fig:fig2}). 
These periodicities differ from those obtained with time-delay variations. 
Here, we note that the data in this case exhibit clear signatures of Rieger periods and QBOs. 

\subsubsection{Tilt scatter around Joy's law}

\begin{figure}
\includegraphics[width=\linewidth]{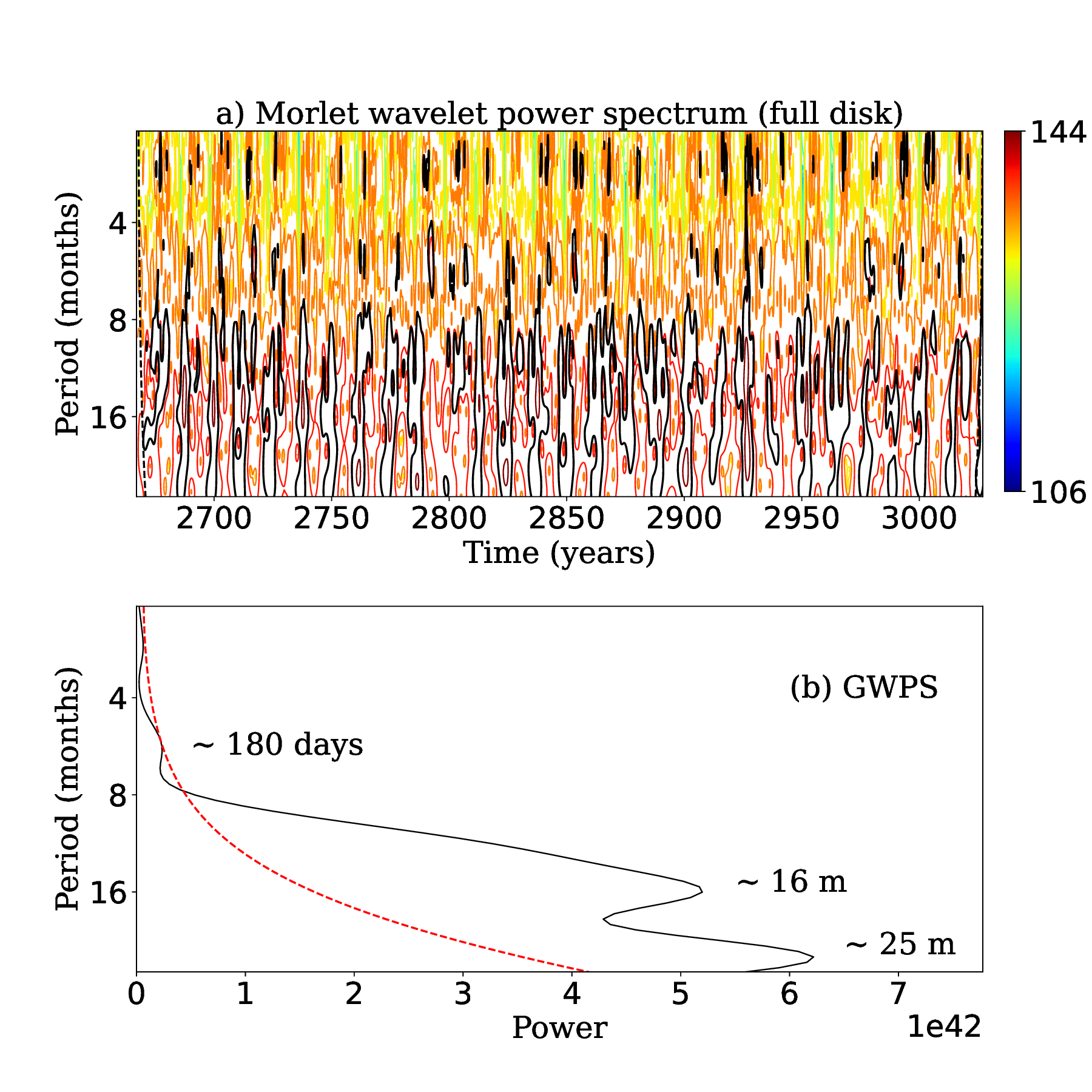}
\includegraphics[width=\linewidth]{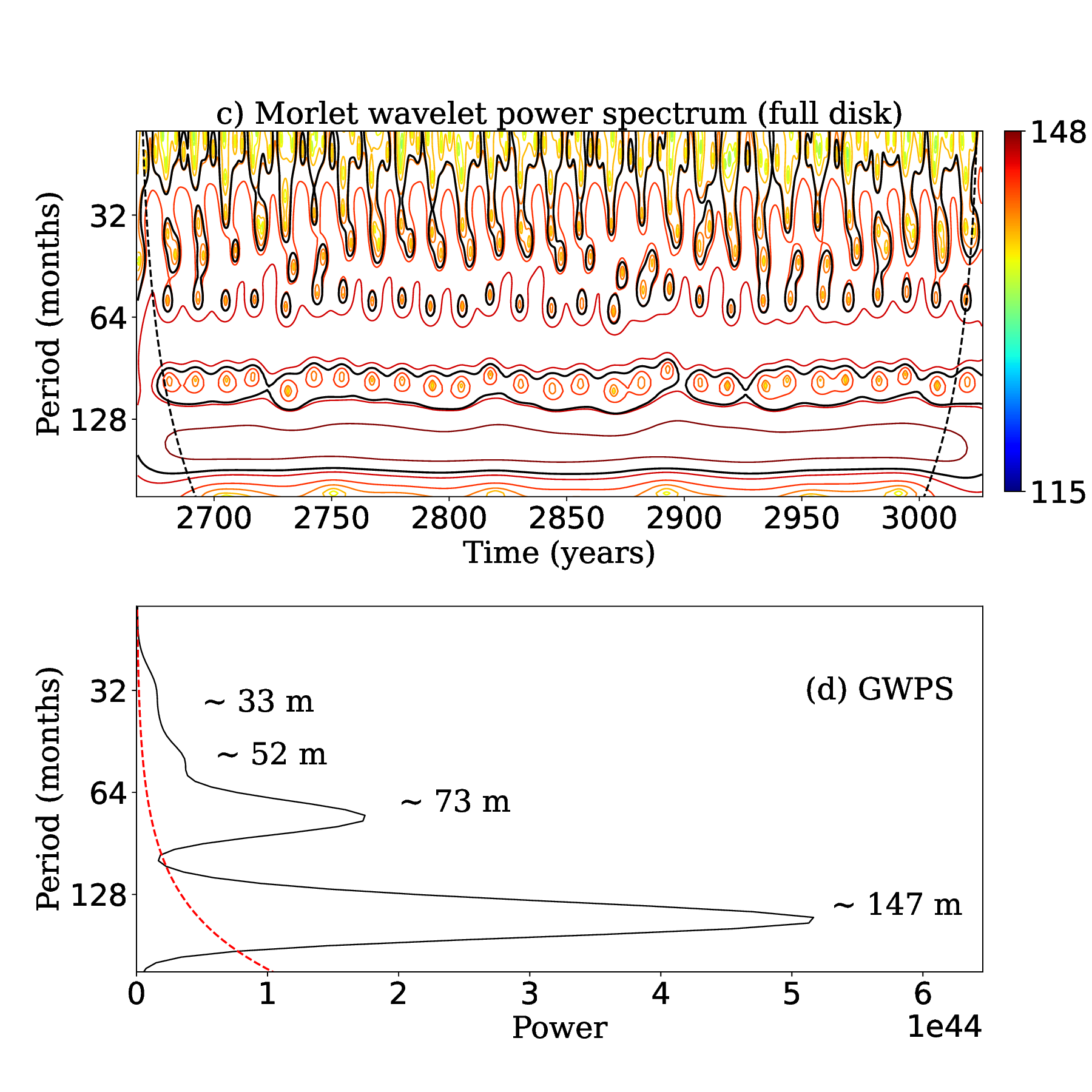}
\caption{Same as \Fig{fig:fig1}, but fluctuation only in the tilt around Joy's law with $\sigma_\delta = 15^{\circ}$ with fixed time delay and flux.}
\label{fig:fig3}
\end{figure}

In this case, we consider tilt scatter $(\sigma_\delta)$ of $15^{\circ}$ around Joy's law inspired by observations \citep{Sreedevi24} while keeping the magnetic flux and time delay fixed. 
The wavelet spectrum analysis of the data in this case reveals periodicities of $\sim$ 180 days, 16, 25, 33, 52, 73, and 147 months (see \Fig{fig:fig3}). This result also shows Rieger-type period signal and QBOs.

In all of the above cases, flux variation and tilt scatter-driven data show Rieger-type periods and QBOs; however, time delay case do not show Rieger periods with 95\% significance level, but show only QBOs. 
Moreover, the occurrence probability of Rieger-type periods is higher in the flux variation case, and the occurrence probability of QBOs is higher in the tilt-scatter case.
Thus, this result suggests that in the \bl\ dynamo, for Rieger-type periodicity, fluctuation in the BMR flux plays a significant role, and for QBOs, tilt scatter plays a crucial role compared to the time delay.
Next, we will see the combined effect of all these
fluctuations in BMR parameters for the generation of QBOs and Rieger-type periodicities.

\subsubsection{Combined fluctuations in \bl\ process}
\begin{figure}
\includegraphics[width=\linewidth]{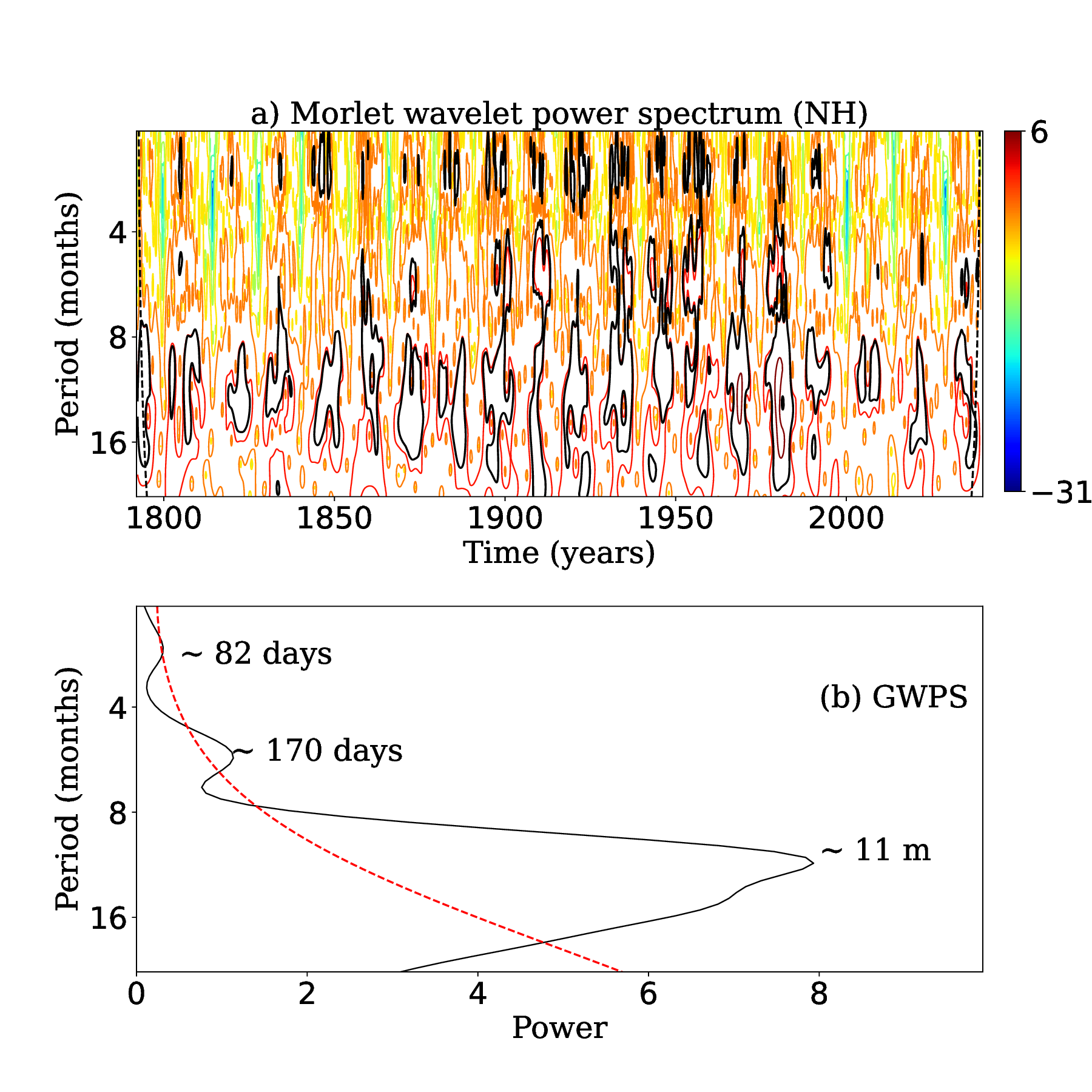}
\includegraphics[width=\linewidth]{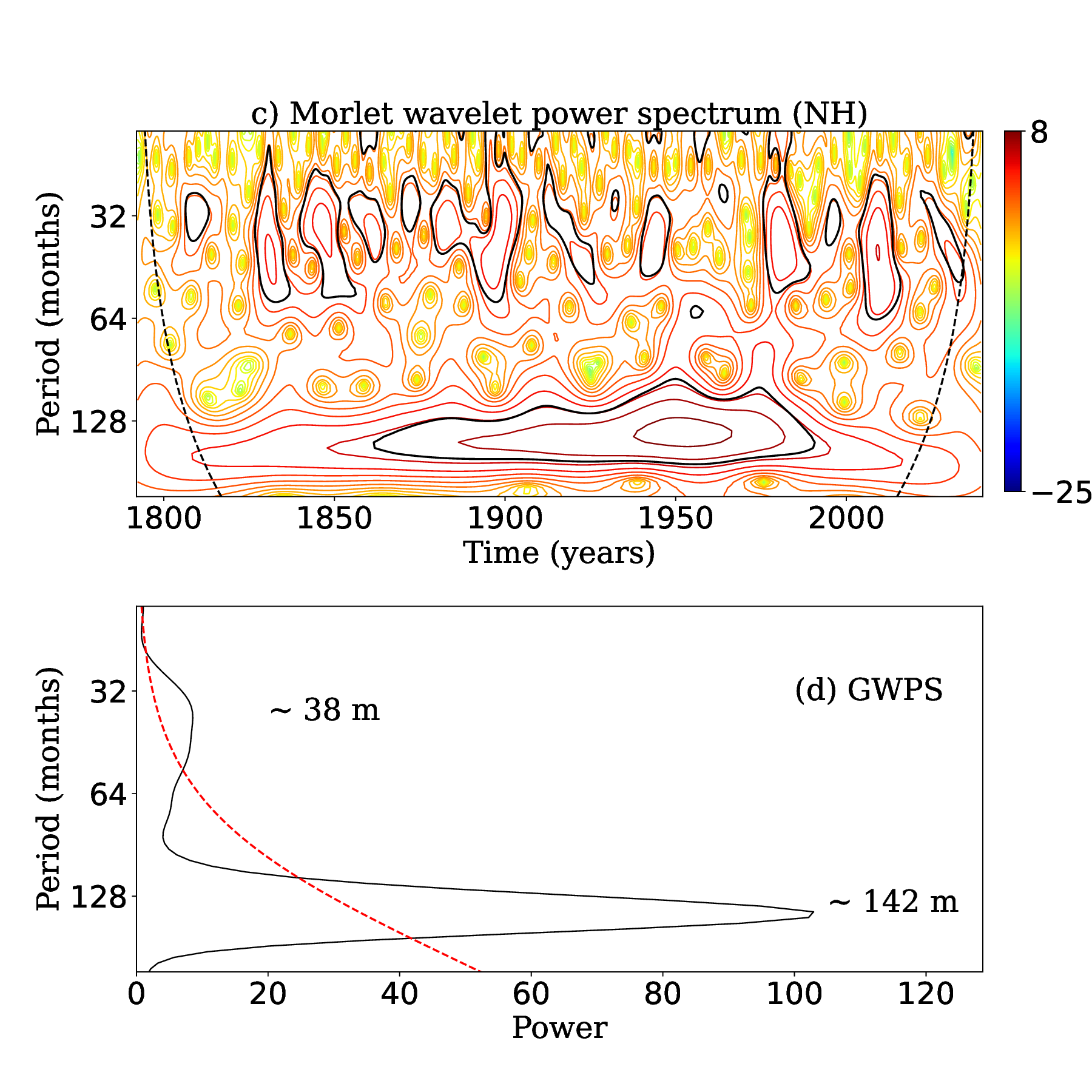}
\caption{The figure shows the dynamical behavior of various periodicities found for the northern hemisphere using wavelet power spectra for the monthly sunspot number time series data obtained from the model with fluctuations in all the parameters of  \bl\ process with tilt scatter $\sigma_\delta = 15^{\circ}$. 
(a) Morlet wavelet spectra to identify short-term periods (Rieger-type).
(b) Global power spectra for short-term periods (Rieger-type). The red dotted line represents the 95\% confidence level. (c) Similar to panel (a) but for the QBOs and long-range periods including the cycle corresponding to the 11-year solar cycle. (d) Similar to panel (b) but for the study of QBOs and long periods.
}
\label{fig:fig4}
\end{figure}

\begin{figure}
\includegraphics[width=\linewidth]{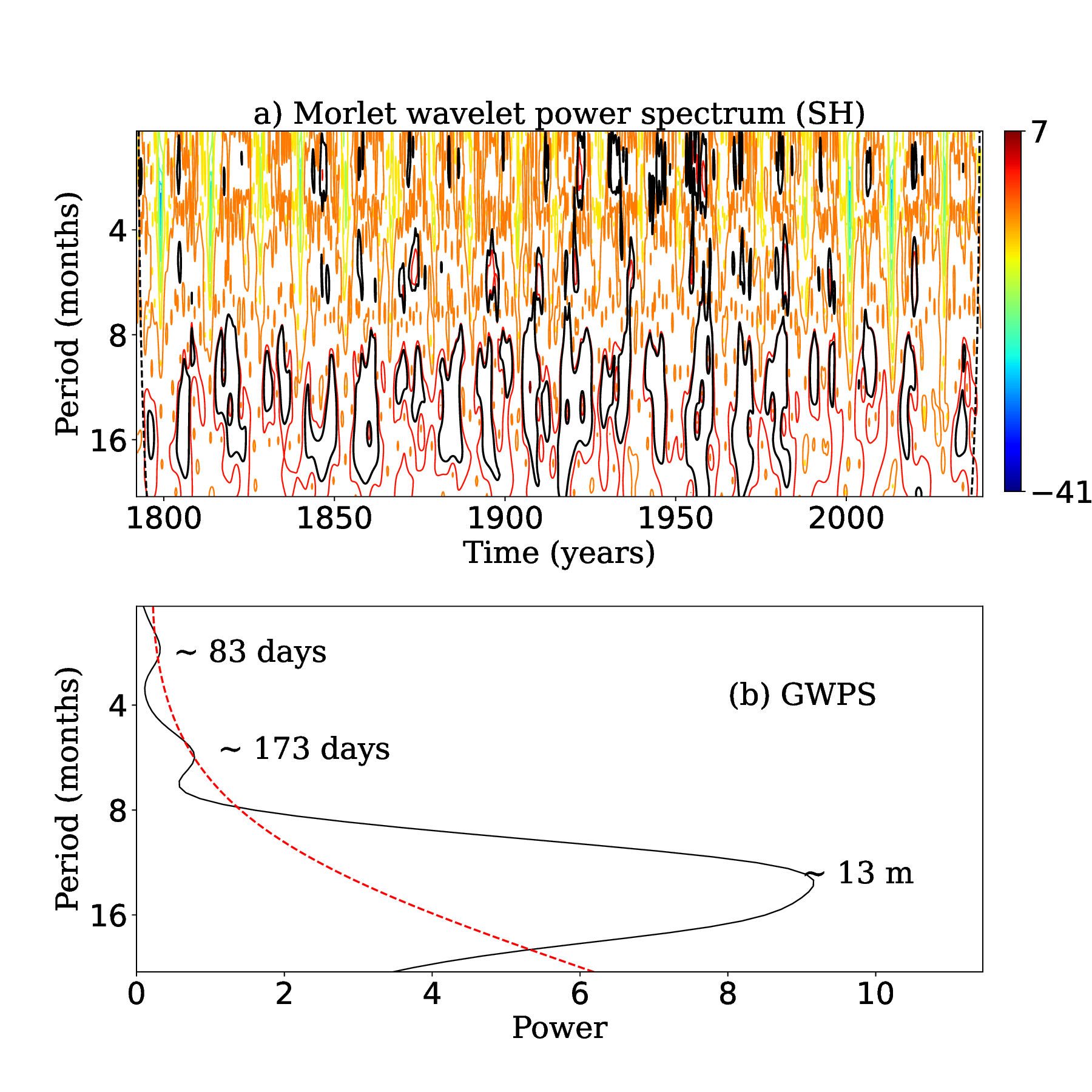}
\includegraphics[width=\linewidth]{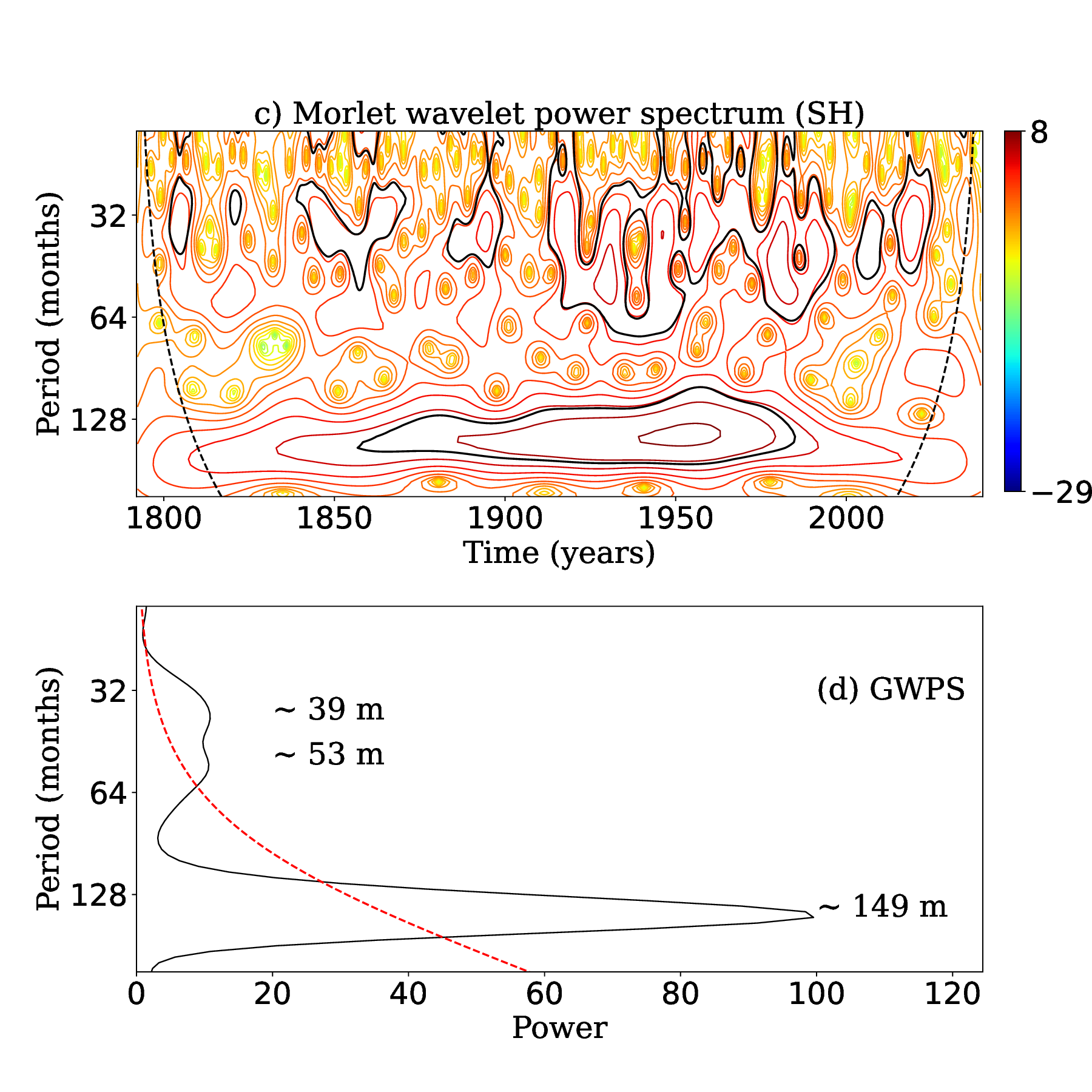}
\caption{Similar to \Fig{fig:fig4}, but for the southern hemisphere.}
\label{fig:fig5}
\end{figure}

\begin{figure}
\includegraphics[width=\linewidth]{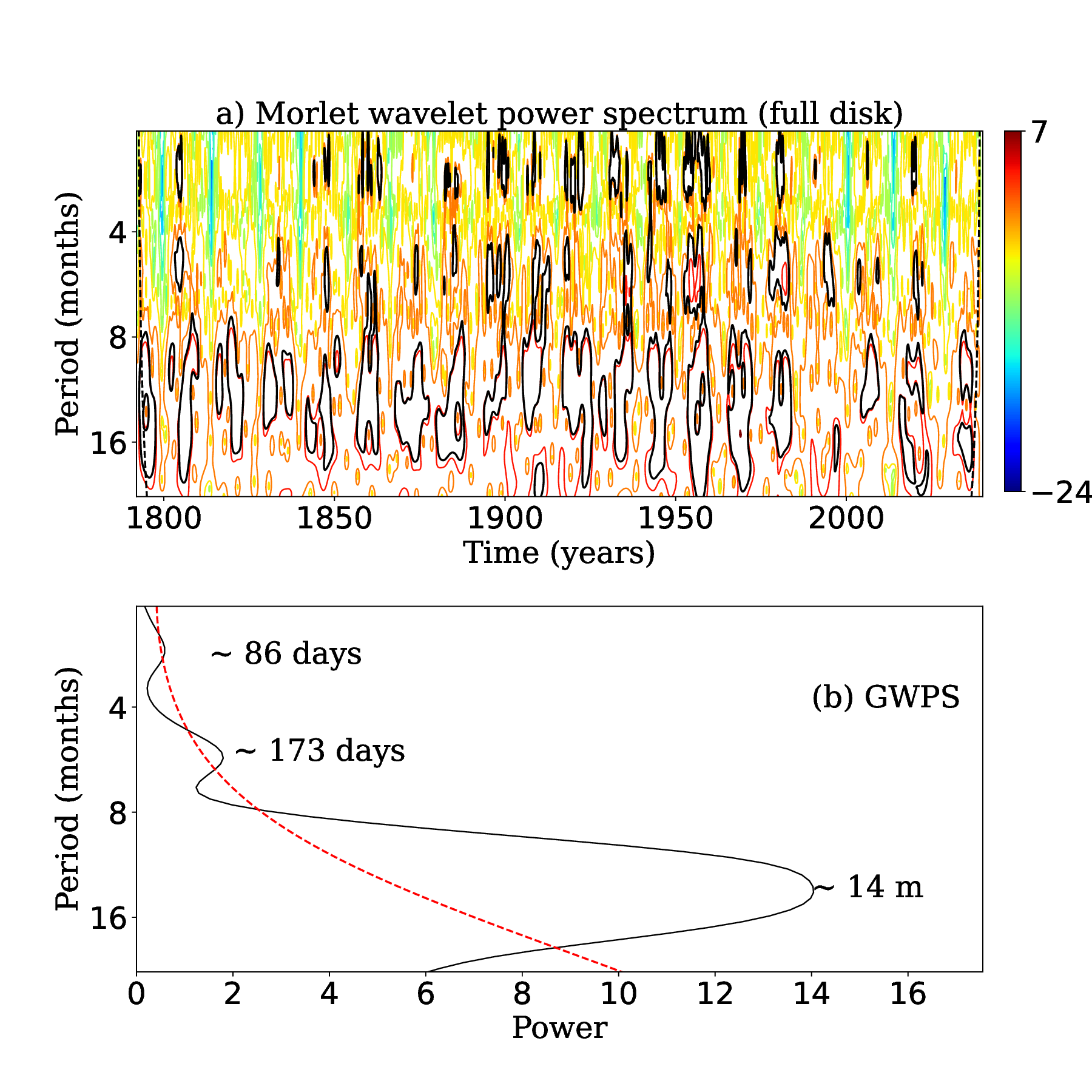}
\includegraphics[width=\linewidth]{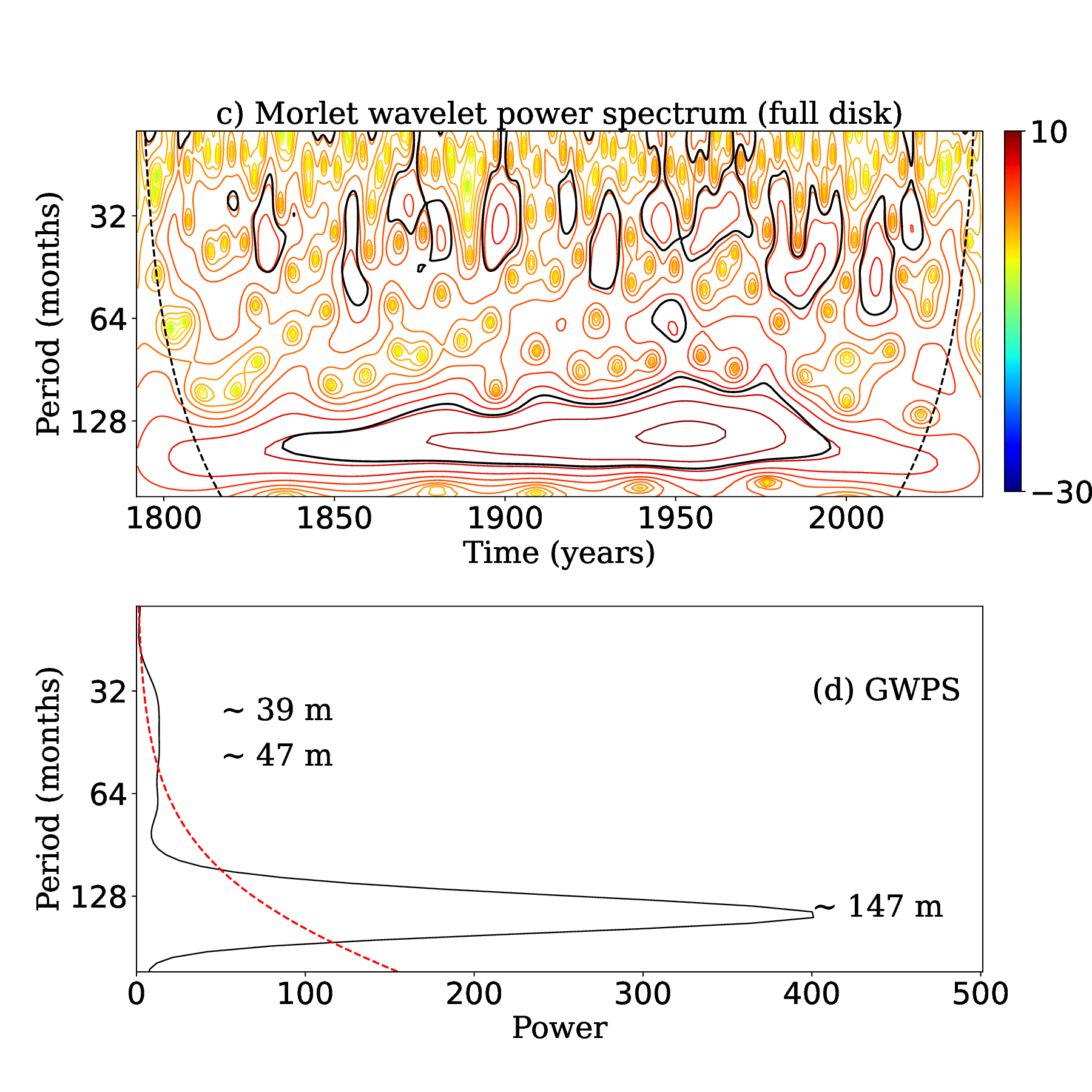}
\caption{Similar to \Fig{fig:fig4}, but for the full-disk data.}
\label{fig:fig6}
\end{figure}

Now, we consider the fluctuations in all  parameters of the \bl\ process, namely the tilt scatter around Joy's law, variation in flux, latitude, and time delay.  In contrast to the previous cases, here we analyse the hemispheric data in addition to the full-disk data. 
For the northern hemisphere, we find $\sim$ 82 and 170 days, 11, 38, and 142 months periodicities (\Fig{fig:fig4}), while for the southern hemisphere, we find $\sim$ 83 and 173 days, 13, 39, 53, and 149 month periodicities (\Fig{fig:fig5}).
Next, we analyze the periods for full-disk data, and we find $\sim$ 86 and 173 days, and 14, 39, 47, and 147-month periodicities in both Morlet wavelet spectra and GWPS (\Fig{fig:fig6}) with 95\% significance level.

We recall that observational studies have reported Rieger-type periodicities in solar magnetic and proxy data, such as $\sim$ 60--85 days \citep{Joshi06}, 154 days \citep{Rieger}, 160 days \citep{B02}, 155 days \citep{Cho14}, 155–-160 days \citep{Za10}, 137–-283 days \citep{RB22}, and 140 –- 158 days \citep{Chowdhury24}. 
Our model also reproduced similar Rieger-type periodicities in the range of $\sim$ 70–-180 days (see \Figss{fig:fig4}{fig:fig6})(a and b).
Furthermore, our model reproduced QBOs 
with periods ranging from $\sim$ 1 to 4.5 years (\Figss{fig:fig4}{fig:fig6})(c and d) which is in close agreement with the observed range of $\sim$ 1 to 4 years often found in various datasets \citep{Baz14, RB22}.
We note that in our analysis, the Rieger-type periods and QBOs contours with 95\% confidence level are not continuous and are confined primarily in the phase of solar maxima.
Thus, the model results suggest that the Rieger-type periodicity and QBOs are intermittent, and the occurrence probability is high during the solar maximum phase, consistent with observations \citep{B02, BO11, KJ23}. 
Moreover, \citet{GE17, GE21} studies show that Rieger-type periods and QBOs exhibit north-south asymmetry.
Our model also reproduces different periodicity in Rieger-type periods and QBOs in both hemispheres (\Tab{table1}). 
The reason for getting QBOs and Rieger-type periods in \bl\ dynamo is the stochastic fluctuations and non-linearity due to individual parameters (time delay, flux, latitude, tilt) in the \bl\ process.  
Studies reveal that stochastic interactions in complex systems induce transient periodic or quasi-periodic behaviors \citep{GS96, Guo17}. 
Moreover, \citet{OCK13} study suggests that the stochastic fluctuation in the \bl\ process with correlation times comparable to the solar rotation period can induce variability on shorter timescales.

Additionally, we see the effect of tilt scatter around Joy's law on Rieger-type periods and QBOs. 
We start increasing the tilt scatter ($\sigma_\delta$) from $15^{\circ}$ to $25^{\circ}$ and see the effect on periodicities. 
We do not get significant periodic variations in all cases, and the Rieger-type periodicity and QBOs are unaffected by the variation of tilt scatter. 
This behavior arises because, as the tilt scatter increases, the amplitude of polar field fluctuations becomes larger; thus, the strength of the next solar cycle increases; however, the cycle periodicity remains largely unaffected \citep{Ji14, KM17, BKK23}. The periodicity of the solar cycle is primarily determined by the meridional flow, diffusivity, and $\alpha$ effect \citep{KC11}. 
Therefore, we do not see any significant changes in QBOs and Rieger-type periods with the tilt scatter. 

\subsection{Effect of dynamo supercriticality}
\label{sec:sup}
\begin{figure}
\includegraphics[width=\linewidth]{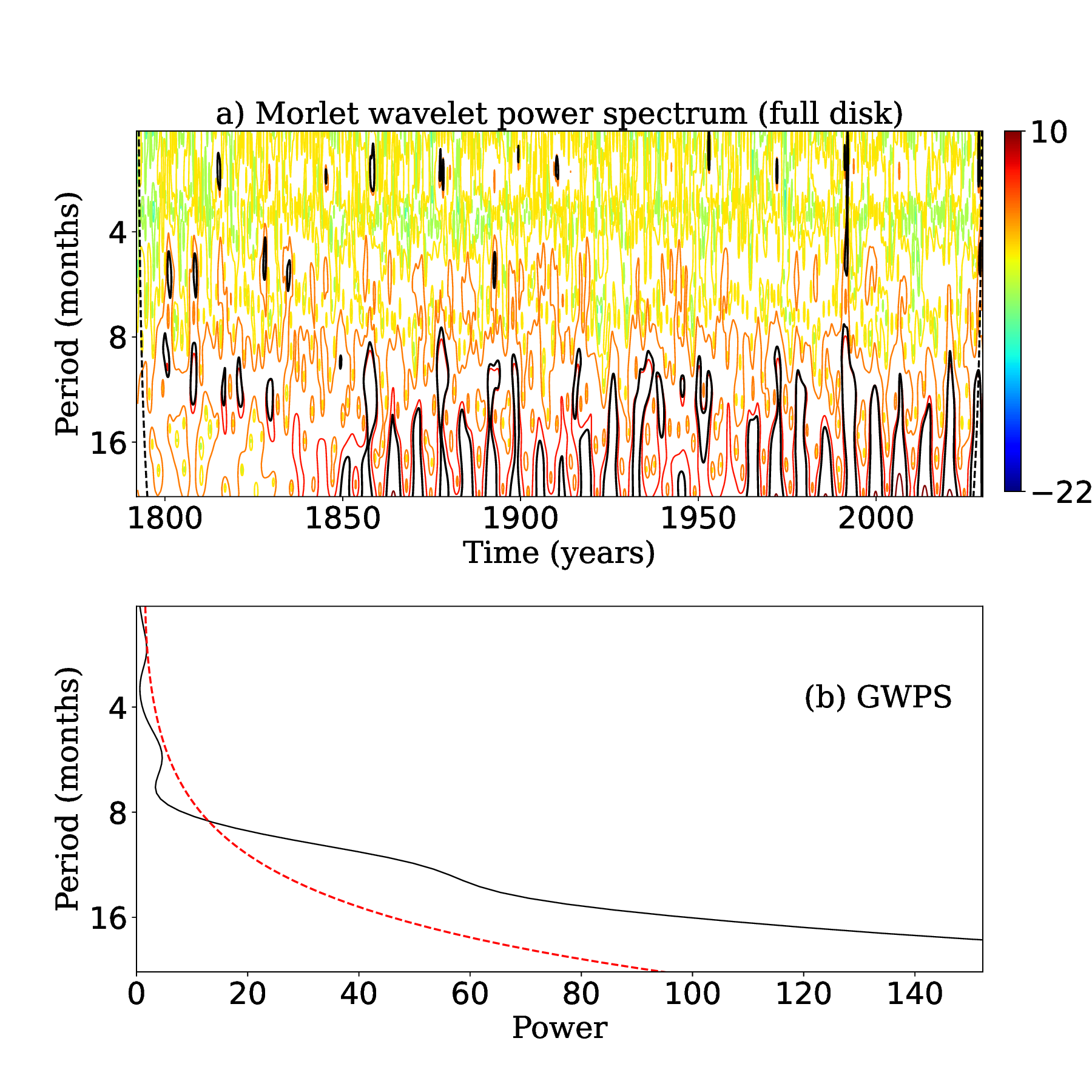}
\includegraphics[width=\linewidth]{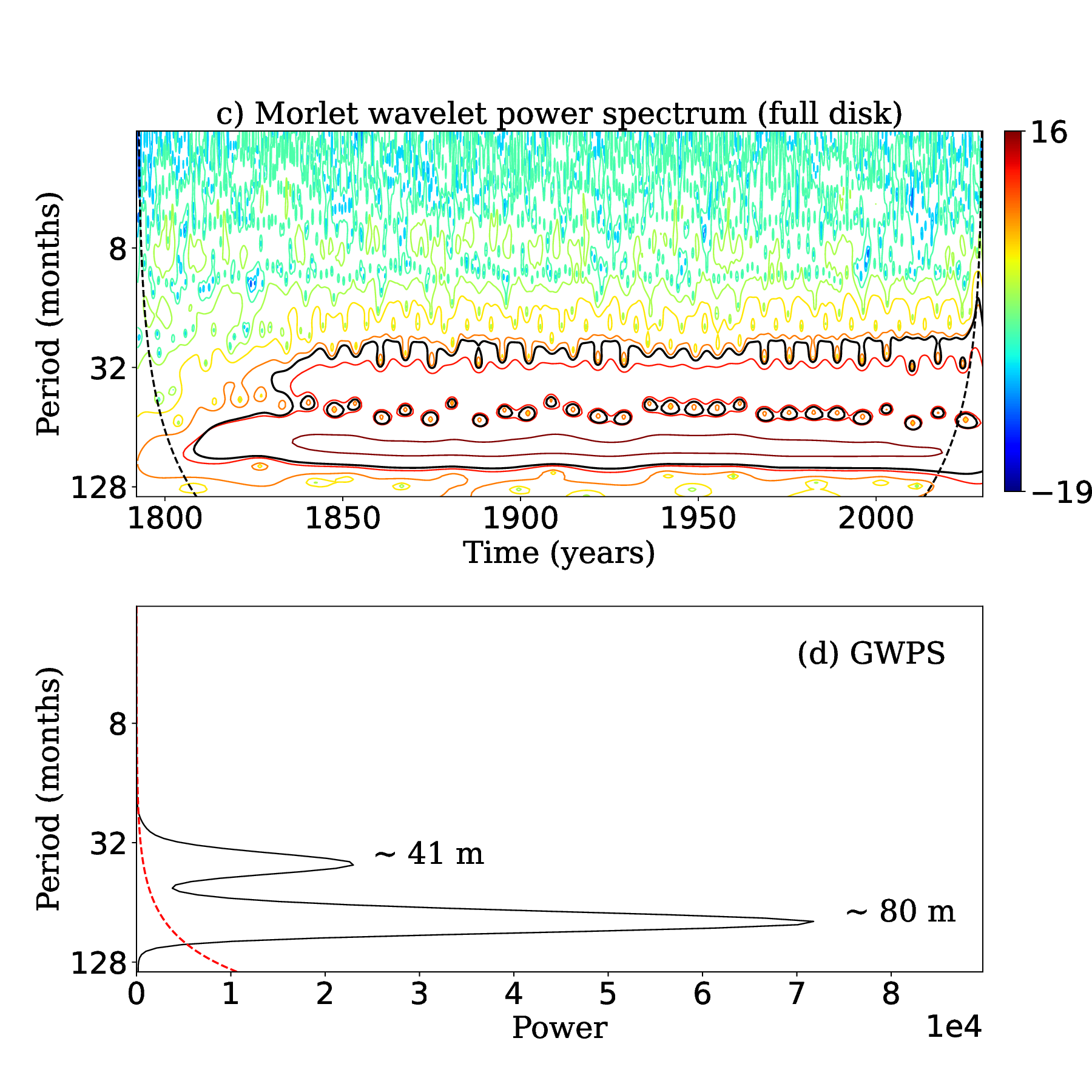}

\caption{Same as the \Fig{fig:fig4}, but at highly supercritical regime of the dynamo with $\Phi_0 = 4$.}
\label{fig:sup}
\end{figure}

So far, our model is operated at weakly supercritical regime. Here, we want to explore the effect of dynamo supercriticality on the properties of QBO and Rieger-type oscillations.  
To increase the supercriticality of the dynamo, we increase the value of $\Phi_0$ (i.e., boost the BMR flux at higher values) from 2 to 4 and keep the other parameters fixed. We note that increasing the value of $\Phi_0$ makes the generation of the polar field efficient. Thus, this   $\Phi_0$ parameter acts in a similar manner as the strength of $\alpha$ in 2D kinematic dynamo models and relates to the dynamo number \citep{MT16, Kar20}.
We find that when $\Phi_0$ is small, i.e., the dynamo is weakly supercritical, we find periodicities of $\sim$ 86 and 173 days, 39, 47, and 147 months in full-disk data,  which correspond to Rieger-type periodicity and QBOs as observed in different observational data \citep{Baz14}.
However,
as the dynamo supercriticality increases, the occurrence probability of Rieger-type periodicity decreases
and when the values of $\Phi_0$ are considerably high, dynamo operates in a highly supercritical regime; we find only QBOs-like periodicities with 95\% significance level (see \Fig{fig:sup}).
The reason is that, as the supercriticality of the dynamo increases, the effect of fluctuations on the magnetic field is suppressed due to increasing influence of the nonlinearity (\citet{kumar21b} or see, Fig. 16 of \citet{Karak23}).
As a result, the \bl\ dynamo in the supercritical regime fails to produce Rieger-type short-term periods (see \Fig{fig:sup}(a) and (b) and \Tab{table1}). 
However, in this regime, the dynamo still produces QBO features (\Fig{fig:sup}(c) and (d)).

Next, our analysis shows $\sim$ 41 and 80 month periods in a highly supercritical dynamo, as shown in \Fig{fig:sup}(c) and (d). 
Here, we note that the 80-month periodicity corresponds to the dominant/regular period (11-year for the sun), which is reduced from 12.25 years at $\Phi_0 = 2$. This reduction in the length of the regular cycle is due to the operation of dynamo at highly supercritical regime. It was already known that increasing the dynamo supercriticality, the cycle duration decreases due to the efficient generation of poloidal field  \citep{Noyes84b}; also see Table~2 of \citet{kumar21b}. 
Furthermore, in our dynamo simulations we observe that with increasing dynamo supercriticality, the regular/dominant cycle period decreases at a faster rate than the other short-term periodicities (\Tab{table1}).
The reason is that the fundamental periods of the solar cycle are primarily governed by the combined effect of the efficiency of poloidal field generation ($\Phi_0$ in the present case) and large-scale flows \citep{DC99, Kar10}, which is unchanged here. 
With the increase of $\Phi_0$, the polar field build-up rate increases; thus, polarity reversal occurs earlier than usual \citep{Cha20, GBK23}. This results in a decrease in the periods of the regular cycle. 
However, the short-term periods are determined by the coherence time of stochastic fluctuations, which remains constant in our model.
Thus, the QBOs and Rieger-type periodicities remain almost unaffected; slight decrease is because of the increase of the dynamo efficiency.

\begin{table}
\centering
\caption{Short-term periods (Rieger-type), QBOs and regular cycle corresponding to 11-year solar cycle at different amounts of the dynamo supercriticality as measured by the $\Phi_0$ value.
}

\begin{tabular}{llclcl} 
\cline{1-4}
$\Phi_0$  & Rieger-type & QBOs & Regular cycle \\
value & Periods (days) & Periods (month) & Periods (years)\\
\cline{1-4}
2 & $\sim$ 86, 173 & $\sim$ 14, 39, 47 & ~~~ 12.25   \\
    
\cline{1-4}

3 & $\sim$ 84, 165 &  $\sim$ 10, 29, 46 & ~~~ 8.67   \\
\cline{1-4}
3.5 & $\sim$ 82, 164  & $\sim$ 13.6, 43 &  ~~~ 6.92  \\
\cline{1-4}
4	& ~~$---$ & $\sim$ 41 &  ~~~ 6.67  \\
\cline{1-4}

\end{tabular}
\label{table1}
\end{table}

\section{Conclusions}
\label{sec:con} 

In this study, we explore QBO and Rieger-type periods and their physical mechanism in the \bl\ dynamo. For this, we use a 3D \bl\ dynamo model (STABLE), which captures the realistic properties of the BMRs and closely connects the model with observed solar magnetic features. For the first time, we show that the \bl\ dynamo can produce Rieger-type periods and QBOs as seen in the different observational data \citep{Rieger, Baz14}.
In particular, we investigate the role of individual 
\bl\ parameters
relating to BMR properties for Rieger-type periodicity and QBOs.
We find that fluctuations in all individual 
\bl\ parameters,
namely the distributions of the BMR time delay, flux, tilt angle, and latitude 
can successfully reproduce the QBOs of
periodicities between $\sim$ 1 to 4.5 years and Rieger-type periods with 95\% significance level 
except for the case of time delay fluctuations, for which we observed only QBOs.
Moreover, when the fluctuations in all the parameters of the \bl\ process act together, the probability of occurrence of QBOs and Rieger periods increases significantly due to the interaction of different stochastic fluctuations in the \bl\ process \citep{PK12, SC18, CS19}.
Next, we analyze the sensitivity of the tilt scatter on the QBOs and Rieger periodicities. We find that the varying tilt scatter does not significantly affect the QBOs and Rieger-type periodicity and appearance. 

We note that \citet{IF19} could not find  Rieger-type periods or QBOs in their 2D \bl\ dynamo model. 
A possible explanation could be that the \bl\ process is inherently 3D, and in the 2D model, this is captured by a single term $\alpha$ through which all complex fluctuating properties of the BMR may not be adequately captured, which are required to reproduce such mid-term periodicities. 

Finally, we investigate the effect of dynamo supercriticality on these periodicities. 
We find that when the dynamo operates near the critical region, the \bl\ dynamo reproduces Rieger-type periodicities and QBOs successfully. 
However, as we increase the supercriticality of the dynamo, the probability of occurrence of Rieger-type periodicity decreases, and in a highly supercritical regime, the dynamo fails to reproduce the Rieger-type periods, while QBOs are still produced with the significance level 95\%.
This result
presents independent evidence for the suggestion that the solar dynamo is not highly supercritical \citep[e.g.,][]{V23,  Ghosh24, Wavhal25}.
Furthermore, we find a decrease in the periodicity of the Rieger-type, QBOs, and regular cycle with the dynamo supercriticality. This result supports the observational findings that weaker cycles show longer Rieger-type periodicities $\sim$ (185--195 days), whereas stronger cycles have shorter ones $\sim$ (155--165 days) \citep{GE17}.
In conclusion, our 3D \bl\ dynamo model including stochastic fluctuations in the BMR parameters successfully reproduces QBOs and Rieger-type periodicities along with their intermittent and asymmetric properties--the key observational features of the solar cycle. 

\begin{acknowledgements}

The authors thank the anonymous referee for offering constructive comments that helped to improve the quality of the manuscript to a large extent. The authors also acknowledge the computational support and the resources provided by the IIA computing facility.
B.B.K. acknowledges the financial support from the Indian Space Research Organization (project no. ISRO/RES/RAC-S/IITBHU/2024-25)

\end{acknowledgements}

%
%
\bibliographystyle{aa}
\bibliography{references}

\end{document}